\documentclass[preprint,12pt]{elsarticle}

\usepackage{amsmath,amssymb,amsfonts}
\usepackage{tabularx}
\usepackage{graphicx}
\usepackage{caption}
\usepackage{float}
\usepackage{booktabs}
\usepackage[utf8]{inputenc}
\usepackage{xcolor}
\usepackage{soul}
\usepackage{listings}
\usepackage{threeparttable}
\usepackage{textcomp}
\usepackage{mathrsfs}
\usepackage[polish,english]{babel}
\usepackage[colorlinks,citecolor=red,urlcolor=blue,bookmarks=false,hypertexnames=true]{hyperref}
\usepackage[version=4]{mhchem}

\journal{Acta Materialia}

\begin{document}
\setlength{\emergencystretch}{3em}

\begin{frontmatter}
	
	\title{High-throughput thermodynamic screening of oxide-scale adhesion across the CoCrFeMnNiAl high-entropy alloys}
	
	\author[a]{Dennis Boakye}
	\author[a]{Chuang Deng \corref{b}}
	\cortext[b]{Corresponding author}
	\ead{chuang.deng@umanitoba.ca}
	
	\affiliation[a]{organization={Department of Mechanical Engineering, University of Manitoba},
		addressline={66 Chancellors Cir},
		city={Winnipeg},
		postcode={R3T 2N2},
		state={Manitoba},
		country={Canada}}
	
	\begin{abstract}
		One significant benefit of reactive element (RE) additions is the colossal improvement in oxide-scale retention during high-temperature oxidation. Selecting optimal RE dopants in high-entropy alloys remains empirical because the relevant thermodynamic landscape is inaccessible to first-principles at the required compositional resolution. Here we apply the macroscopic atom model, coupled with McLean isotherm and Guttmann models, to screen adhesion across nine CoCrFeMnNiAl sub-families at \ce{Cr2O3} and \ce{Al2O3} interfaces, ranking five REs (Hf, Y, Zr, La, Ti) for segregation, adhesion enhancement, and sulfur displacement. The screening reveals an oxide-dependent ranking inversion, with Hf dominating at \ce{Cr2O3} and La dominating at \ce{Al2O3}, driven by the interplay between RE--O and RE--matrix interaction enthalpies. Mn-containing alloys exhibit intrinsic sulfur resistance, consistent with their experimentally observed oxidation characteristics. A sulfur immunity phase diagram identifies compositions with Al~+~Mn~$\gtrsim$~25~at\% as thermodynamically immune to S-induced adhesion loss. All crossover concentrations collapse onto a universal exponential governed by the segregation enthalpy difference, providing a transferable design rule. Inverse design identifies \ce{Co16Cr16Fe16Ni16Al35} as the optimal S-immune composition with $W_\text{sep} = 5.95$~J/m$^2$ without RE doping.
	\end{abstract}
	
	
\end{frontmatter}

%% =====================================================================
\section{Introduction}
\label{sec:intro}
%% =====================================================================

High-entropy alloys (HEAs) based on the CoCrFeMnNiAl elemental palette have emerged as promising candidates for structural applications in aggressive high-temperature environments~\cite{cantor2004microstructural,yeh2004nanostructured,miracle2017critical}. The defining feature of these alloys, usually near-equiatomic mixing of four or more principal elements stabilized by high configurational entropy, produces exceptional combinations of strength, ductility, and corrosion resistance that can surpass conventional superalloys~\cite{miracle2017critical,lu2020hf,liu2023comparative}. However, the long-term performance of any high-temperature alloy is ultimately limited not by bulk mechanical properties but by the integrity of its protective oxide scale~\cite{birks2006introduction,holcomb2015oxidation,fontana1967corrosion}. When this scale spalls, the alloy loses its protective barrier and undergoes accelerated degradation. Understanding and predicting the adhesion of oxide scales to HEA substrates is therefore a central challenge for the deployment of these materials.

Extensive experimental work has established the oxidation phenomenology of key HEA families. For the CoCrFeMnNi (Cantor) system, Laplanche et al.~\cite{laplanche2016oxidation} showed that Mn-rich oxides ($\alpha$-\ce{Mn2O3} at 600--800$^\circ$C, \ce{Mn3O4} at 900$^\circ$C) dominate the scale, with an oxidation activation energy of $130 \pm 10$~kJ/mol controlled by Mn diffusion through the oxide rather than sluggish diffusion in the alloy matrix~\cite{tsai2013sluggish}. Dehury et al.~\cite{dehury2025experimental} extended these observations to 30-day exposures at 1000$^\circ$C, demonstrating that Mn-free CoCrFeNi forms a more continuous, better-adherent oxide with an oxidation rate constant one order of magnitude lower than that of the Cantor alloy, while Cr-free CoFeMnNi undergoes catastrophic oxidation with phase decomposition. Holcomb et al.~\cite{holcomb2015oxidation} confirmed this hierarchy across eight CoCrFeMnNi variants, finding that low-Mn, high-Cr compositions exhibited the best scale retention. For Al-containing systems, Butler and Weaver~\cite{butler2016oxidation} showed that $\geq$20~at\% Al promotes continuous protective \ce{Al2O3}, while Lu et al.~\cite{lu2020hf} demonstrated that co-doping AlCoCrFeNi with only 0.02~at\% each of Y and Hf produces a uniform $\alpha$-\ce{Al2O3} scale just 4.6~$\mu$m thick after 1000~h at 1100$^\circ$C. Comprehensive reviews~\cite{veselkov2021high} have concluded that Al and Cr are the most important elements for protective oxide formation, while Mn degrades oxidation resistance through porous, poorly adhering scale phases.

Scale adhesion is controlled by the chemistry of the oxide--metal interface, governed by the segregation of trace impurities and dopants~\cite{pint1996experimental,fritscher2023reactive}. Sulfur at bulk concentrations below 30~ppm can accumulate at oxide--metal interfaces to levels sufficient to weaken adhesion and promote spallation~\cite{hou2008segregation,jayne1993sulfur}, as documented in Ni-based superalloys~\cite{smialek1987effect,evans2011oxidation} and increasingly in HEA systems~\cite{holcomb2015oxidation,liu2023comparative}. The addition of REs such as Hf and Y at trace levels counteracts sulfur through competitive interfacial displacement, modification of oxide growth kinetics, and formation of stable RE--O pegs that anchor the scale~\cite{pint1996experimental,moon1989role,pint1995reactive,hou1995effect,jedlinski1993comments}. Chen et al.~\cite{chen2024selective} further demonstrated that selective oxidation of Cr, Co, and Fe in Al$_{0.1}$CrCoFeNi creates a Ni-enriched subsurface layer that hinders repassivation, showing how compositional changes at the interface directly control the integrity of the protective film.

Despite decades of research, RE selection for a given alloy--oxide combination remains largely empirical. With six principal elements, five RE candidates, two oxide types, and continuous composition ranges, the design space is inaccessible to exhaustive experimental exploration. First-principles density functional theory has provided valuable atomistic insight for specific systems~\cite{boakye2024reactive,jiang2008first,boakye2025effect,lan2014effects}, but the computational cost of interface supercell calculations limits coverage to isolated configurations. In our previous paper~\cite{boakye2026rapid}, we extended the macroscopic atom model (MAM)~\cite{niessen1989macroscopic,de1988cohesion} to multicomponent alloys, where we validated the framework against DFT for CoCrFeNi and AlCoCrFeNi systems. Here we apply this framework to screen adhesion across nine sub-families of the CoCrFeMnNiAl system at both \ce{Cr2O3} and \ce{Al2O3} interfaces, ranking five REs for segregation tendency, adhesion enhancement, and sulfur displacement, and systematically comparing the predictions with the experimental oxidation literature.

%% =====================================================================
\section{Computational Methods}
\label{sec:methods}
%% =====================================================================

\subsection{Extended macroscopic atom model}
\label{sec:mam}

The theoretical framework is described in detail in our companion paper~\cite{boakye2026rapid}; here we summarize the essential equations. The work of separation $W_\text{sep}$ is computed from the Dupr\'e relation,
\begin{equation}
	W_\text{sep} = \gamma_\text{HEA} + \gamma_\text{TGO} - \gamma_\text{int},
	\label{eq:dupre}
\end{equation}
where $\gamma_\text{HEA}$ and $\gamma_\text{TGO}$ are the surface energies of the alloy and thermally grown oxide (TGO), and $\gamma_\text{int}$ is the interfacial energy. The surface energy of a multicomponent alloy is evaluated from self-consistent surface fractions $C_i^S$ and corrected molar volumes through an iterative procedure~\cite{boakye2026rapid}:
\begin{equation}
	\gamma_0 = \frac{1}{f_\text{vac} \, c_0} \sum_{i=1}^{n} C_i^S \frac{\Delta H_i^\text{surf}}{V_i^{2/3}},
	\label{eq:surface_energy}
\end{equation}
with $f_\text{vac} = 0.31$, $c_0 = 4.5 \times 10^8$~mol$^{-1/3}$, $\Delta H_i^\text{surf}$ the surface enthalpy, and $V_i$ the corrected molar volume of species $i$. The interfacial energy is obtained from cross-interface pair contact probabilities and Miedema interaction enthalpies $\Delta H_{ij}^\circ$~\cite{de1988cohesion}.

The segregation enthalpy of a solute $A$ at the HEA--TGO interface decomposes into four physically distinct contributions~\cite{boakye2025effect}:
\begin{equation}
	\Delta H_\text{seg}^\text{int}(A) = \frac{1}{3}\left[
	\sum_j C_j^\text{TGO} \Delta H_{Aj}^\circ
	- \sum_{ij} P_{ij}^\text{int} \Delta H_{ij}^\circ
	- \sum_i C_i^\text{HEA} \Delta H_{Ai}^\circ
	\right] + \Gamma_\text{HEA}.
	\label{eq:seg_enthalpy}
\end{equation}
The first term in brackets of Eq.~\eqref{eq:seg_enthalpy} captures the chemical attraction between the solute and the oxide surface, the second term represents the bonding across the clean interface that must be disrupted when the solute replaces a matrix atom, and the third term accounts for the energy cost of removing the solute from the bulk alloy; the last term captures the elastic strain energy from the size mismatch between solute and host.

The equilibrium interfacial coverage follows the McLean isotherm~\cite{mclean1958grain},
\begin{equation}
	\frac{C_A^\text{int}}{1 - C_A^\text{int}} = \frac{c_A^\text{bulk}}{1 - c_A^\text{bulk}} \exp\left(-\frac{\Delta H_\text{seg}^\text{int}}{k_B T}\right),
	\label{eq:mclean}
\end{equation}
and for two competing solutes $A$ and $B$, the crossover concentration at which $A$ displaces $B$ is~\cite{guttmann1976link}
\begin{equation}
	c_A^* = c_B \exp\left(\frac{\Delta H_\text{seg}^A - \Delta H_\text{seg}^B}{k_B T}\right).
	\label{eq:crossover}
\end{equation}

\subsection{Model scope and limitations}
\label{sec:limitations}

Several assumptions underlie the MAM framework. The Miedema interaction enthalpies are derived from binary systems and applied here under a pairwise additivity assumption, which has been validated for surface energy predictions in ternary and quaternary alloys~\cite{de1988cohesion}, but may deviate in concentrated multicomponent environments where higher-order interactions become significant. The McLean isotherm assumes dilute segregation, which is appropriate for trace RE and S additions but may be inaccurate at the high coverages predicted at low temperatures. The model treats the interface as atomically sharp and uses fixed stoichiometric oxide surface compositions (Cr:O = 0.4:0.6; Al:O = 0.4:0.6), neglecting the graded composition profiles and non-stoichiometric defects present in real oxide scales~\cite{laplanche2016oxidation,dehury2025experimental}. Additionally, the MAM computes oxide surface energies by treating the oxide as a pseudo-alloy of its metallic and non-metallic constituents, which overestimates $\gamma_\text{oxide}$ relative to experimental values. Since this enters $W_\text{sep}$ as a constant additive offset for all alloys at a given oxide, all inter-alloy rankings, $\Delta W_\text{sep}$ enhancements, segregation enthalpies, and crossover concentrations are unaffected; the model's predictive strength therefore lies in relative comparisons rather than absolute adhesion energies. Despite these approximations, validation against DFT~\cite{boakye2026rapid} demonstrates that the MAM correctly reproduces segregation hierarchies and adhesion trends, providing a reliable basis for comparative screening.

\subsection{Screening protocol}
\label{sec:protocol}

Nine sub-families of the CoCrFeMnNiAl system were selected for screening (Table~\ref{tab:families}), spanning equiatomic quaternary through senary compositions that cover the major experimentally studied HEA systems. Five RE candidates comprising Hf, Y, Zr, La, and Ti were chosen based on available literature, industrial relevance and the availability of complete Miedema parameter sets~\cite{de1988cohesion,neuhausen2003extension}.

\begin{table}[htbp]
	\centering
	\caption{HEA sub-families screened. $n$ is the number of elements in the alloy. All compositions are equiatomic.}
	\label{tab:families}
	\begin{tabular}{llccc}
		\toprule
		Family & $n$ & Phase & Oxide type \\
		\midrule
		Cantor (CoCrFeMnNi) & 5 & FCC~\cite{dehury2025experimental} & Chromia~\cite{holcomb2015oxidation} \\
		CoCrFeNi            & 4 & FCC~\cite{dehury2025experimental} & Chromia~\cite{holcomb2015oxidation} \\
		AlCoCrFeNi          & 5 & BCC+B2~\cite{butler2016oxidation} & Alumina~\cite{butler2016oxidation} \\
		AlCoCrFeMnNi        & 6 & FCC+BCC~\cite{abdi2023evaluation} & Mixed~\cite{abdi2023evaluation} \\
		CoCrMnNi            & 4 & FCC~\cite{wu2014recovery}         & Chromia \\
		AlCoCrMnNi          & 5 & Unknown                           & Alumina \\
		CoCrFeMn            & 4 & Unknown                           & Chromia \\
		AlCrFeNi            & 4 & B2+A2~\cite{gawel2026tailoring}   & Alumina~\cite{gawel2026tailoring} \\
		CrFeMnNi            & 4 & FCC~\cite{gaber2025optimization}  & Chromia \\
		\bottomrule
	\end{tabular}
\end{table}

For each family, seven quantities were computed at both \ce{Cr2O3} and \ce{Al2O3} interfaces: the baseline $W_\text{sep}$, surface energy, segregation enthalpies of all five REs and S, crossover concentrations for each RE to displace 50~ppm S at 1373~K, adhesion enhancement $\Delta W_\text{sep}$ at 1~at\% RE, McLean isotherm coverages at 100 and 1000~ppm, and a composite RE effectiveness index. For the Cantor alloy, two-dimensional adhesion maps were computed over the (Cr, Fe) composition space at 100~ppm Hf, Zr, and Ti. The entire screening comprising over 500 individual calculations completed in under 3~minutes on a single processor core. All Miedema parameters are taken from de Boer et al.~\cite{de1988cohesion} and Neuhausen and Eichler~\cite{neuhausen2003extension}, and complete tables are provided in the Supplementary Material.

%% =====================================================================
\section{Results and Discussion}
\label{sec:results}
%% =====================================================================

\subsection{Surface energies across HEA families}
\label{sec:surface_energies}

\begin{figure}[!ht]
	\centering
	\includegraphics[width=1\linewidth]{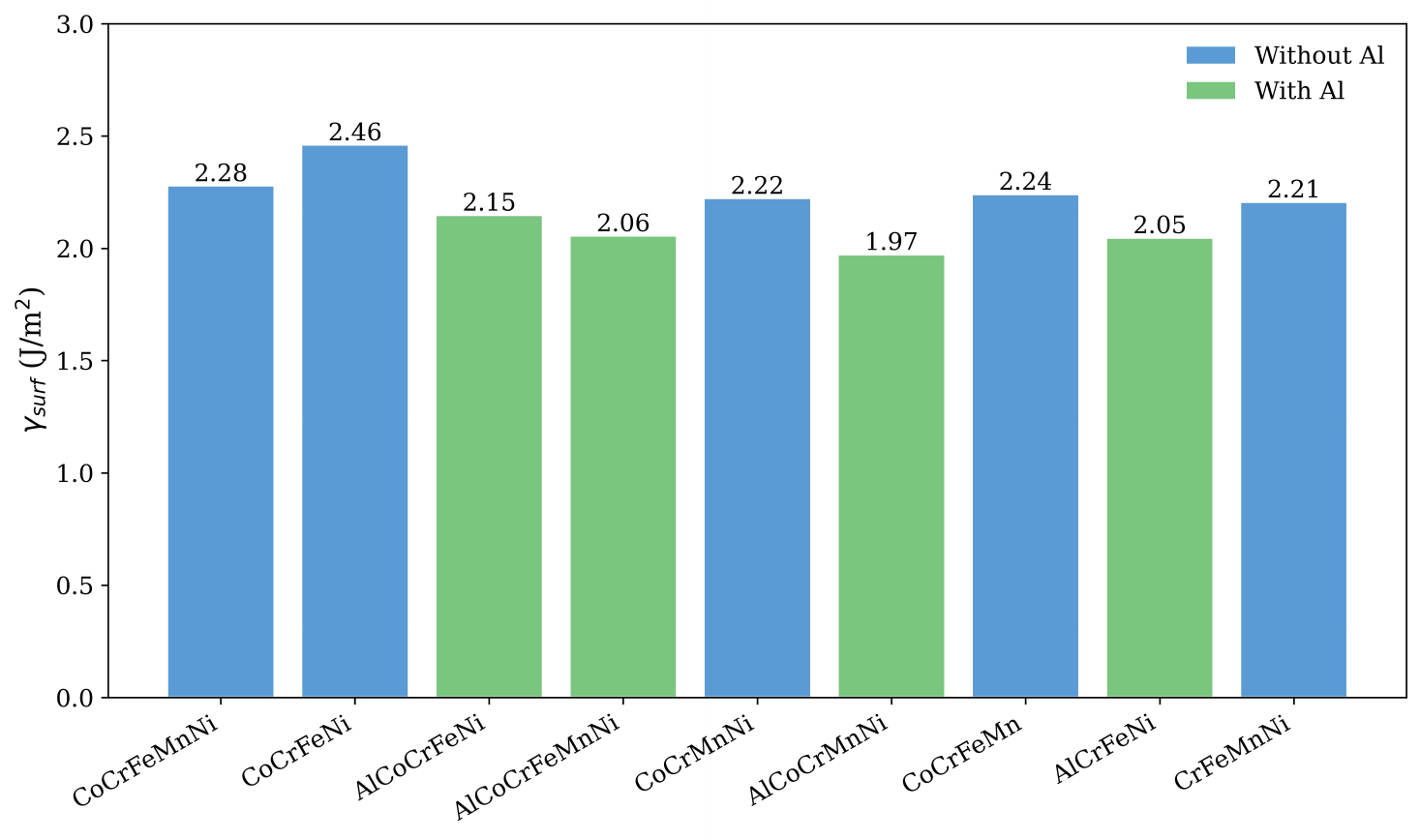}
	\caption{Surface energies of nine HEA sub-families. Al-containing alloys (green) consistently show lower surface energies than Al-free compositions (blue), reflecting the low surface enthalpy of Al (76\,kJ/mol). CoCrFeNi has the highest surface energy (2.46\,J/m$^2$); AlCoCrMnNi the lowest (1.97\,J/m$^2$).}
	\label{fig:surface_energies}
\end{figure}

The computed alloy surface energies range from 1.97~J/m$^2$ (AlCoCrMnNi) to 2.46~J/m$^2$ (CoCrFeNi), with a clear dichotomy between Al-containing and Al-free compositions (Fig.~\ref{fig:surface_energies}). Al-containing alloys consistently exhibit lower surface energies by 0.2--0.3~J/m$^2$, reflecting the low surface enthalpy of Al (76~kJ/mol) compared to Co (127), Cr (121), Fe (129), Ni (121), and Mn (85~kJ/mol)~\cite{de1988cohesion}. This trend is consistent with the predicted preferential surface enrichment of Al by the Gibbs adsorption isotherm. Within each group, Mn-containing variants show intermediate values; for example, the Cantor alloy (2.28~J/m$^2$) lies below CoCrFeNi (2.46~J/m$^2$) due to Mn's lower surface enthalpy. The predicted value for CoCrFeNi agrees well with DFT calculations reporting 2.3--2.6~J/m$^2$ depending on crystallographic orientation~\cite{zhang2022computational}. Because the alloy surface energy enters both the numerator of $W_\text{sep}$ and the interfacial energy term with partially canceling contributions, the cross-interface chemical interactions, rather than the surface energies themselves, emerge as the dominant factor differentiating adhesion across families, as discussed in Section~\ref{sec:baseline_adhesion}.

\subsection{Baseline adhesion and the role of Mn}
\label{sec:baseline_adhesion}

\begin{figure}[tb]
	\centering
	\includegraphics[width=1\linewidth]{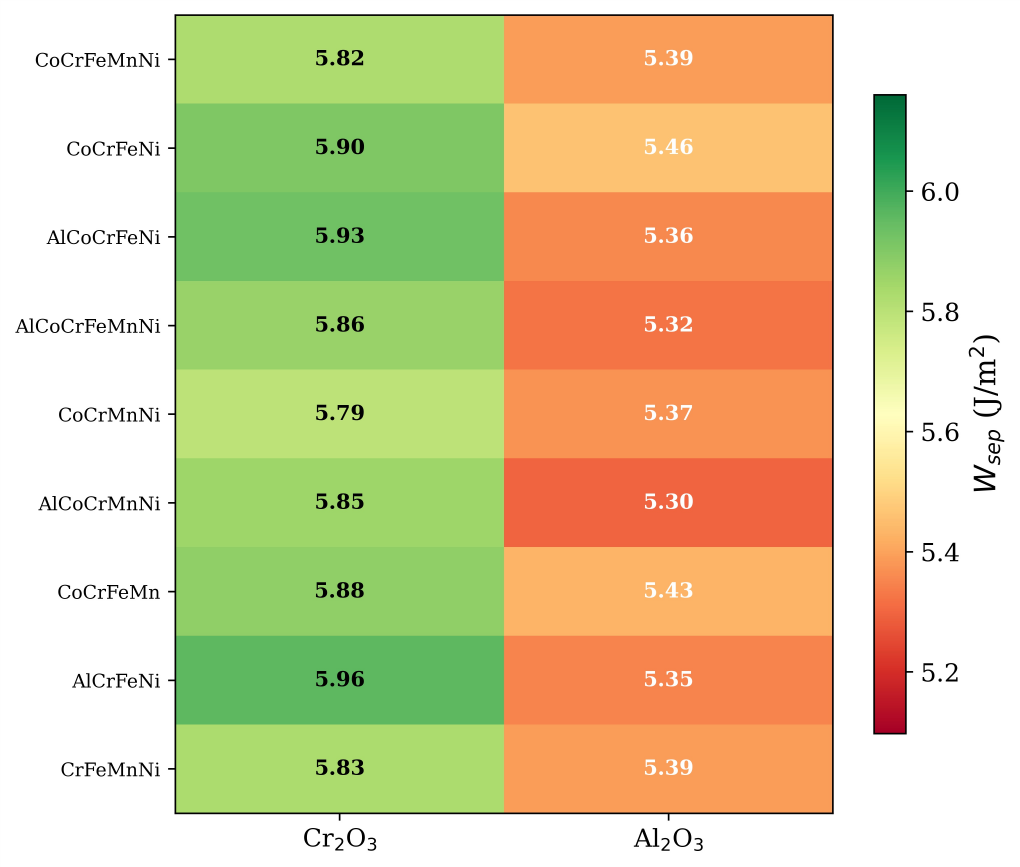}
	\caption{Baseline $W_\text{sep}$ for nine HEA sub-families at \ce{Cr2O3} and \ce{Al2O3} interfaces. Chromia interfaces consistently show higher adhesion. AlCrFeNi produces the highest $W_\text{sep}$ at \ce{Cr2O3} (5.96~J/m$^2$), while AlCoCrMnNi shows the lowest at \ce{Al2O3} (5.24~J/m$^2$).}
	\label{fig:baseline_adhesion}
\end{figure}

Chromia interfaces consistently produce higher $W_\text{sep}$ than alumina across all alloy compositions (Fig.~\ref{fig:baseline_adhesion}), with differences ranging from 0.43~J/m$^2$ (Cantor) to 0.60~J/m$^2$ (AlCrFeNi). This systematic advantage arises primarily from the stronger Cr--O cross-interface bonding ($\Delta H^\circ = -578$~kJ/mol) relative to the Al--O interactions already present within the alumina lattice, consistent with the experimental observation that chromia-forming alloys tend to exhibit superior scale adhesion under moderate oxidation conditions~\cite{pint1995reactive,hou1995effect}.

Two compositional trends emerge from the baseline screening. First, Al increases $W_\text{sep}$ at \ce{Cr2O3} (CoCrFeNi: 5.90 $\rightarrow$ AlCoCrFeNi: 5.93~J/m$^2$) through the strong Al--Cr interaction ($\Delta H_\text{AlCr}^\circ = -87$~kJ/mol), while decreasing it at \ce{Al2O3} (5.46 $\rightarrow$ 5.36~J/m$^2$) because like-element contacts across the interface contribute nothing to the chemical bonding. Second, Mn systematically reduces $W_\text{sep}$ at both interfaces: CoCrFeNi (5.90) $\rightarrow$ Cantor (5.82~J/m$^2$), and AlCoCrFeNi (5.93) $\rightarrow$ AlCoCrFeMnNi (5.87~J/m$^2$). This reduction of 0.06--0.08~J/m$^2$ arises from the weaker Mn--O interaction ($-520$~kJ/mol) relative to Cr--O ($-578$), Co--O ($-302$), and Fe--O ($-413$~kJ/mol).

This Mn-induced adhesion penalty aligns with extensive experimental evidence. Laplanche et al.~\cite{laplanche2016oxidation} reported scale spallation during isothermal oxidation of the Cantor alloy, manifested as small weight losses in TGA curves. More directly, Dehury et al.~\cite{dehury2025experimental} found that after 30 days at 1000$^\circ$C, CoCrFeNi formed a relatively continuous oxide layer with good surface adherence, whereas CoCrFeMnNi exhibited a rough, discontinuous, and significantly spalled scale with Kirkendall pores at the metal--oxide interface. Their measured oxidation rate constant for CoCrFeNi ($8.8 \times 10^{-5}$~(mg/mm$^2$)$^3$/s) was approximately seven times lower than for CoCrFeMnNi ($59 \times 10^{-5}$~(mg/mm$^2$)$^3$/s). The adhesion hierarchy predicted here, CoCrFeNi $>$ CoCrFeMnNi $\gg$ CoFeMnNi, matches the experimental rankings at \ce{Cr2O3} maps.

The mechanistic origin of the Mn effect is also consistent between model and experiment. The model attributes the adhesion reduction to weaker Mn--O cross-interface bonds, while experiments show that Mn diffuses outward through the oxide faster than any other element~\cite{laplanche2016oxidation,dehury2025experimental,holcomb2015oxidation}, forming a Mn-rich outer layer whose growth rate exceeds that of the parent alloy by an order of magnitude. The rapid outward Mn diffusion creates Kirkendall porosity~\cite{dehury2025experimental}, further weakening adhesion through mechanical void-mediated decohesion, thereby amplifying the thermodynamic weakness predicted here.

\subsection{RE segregation ranking and oxide-dependent inversion}
\label{sec:re_ranking}

\begin{figure}[!ht]
	\centering
	\includegraphics[width=0.8\linewidth]{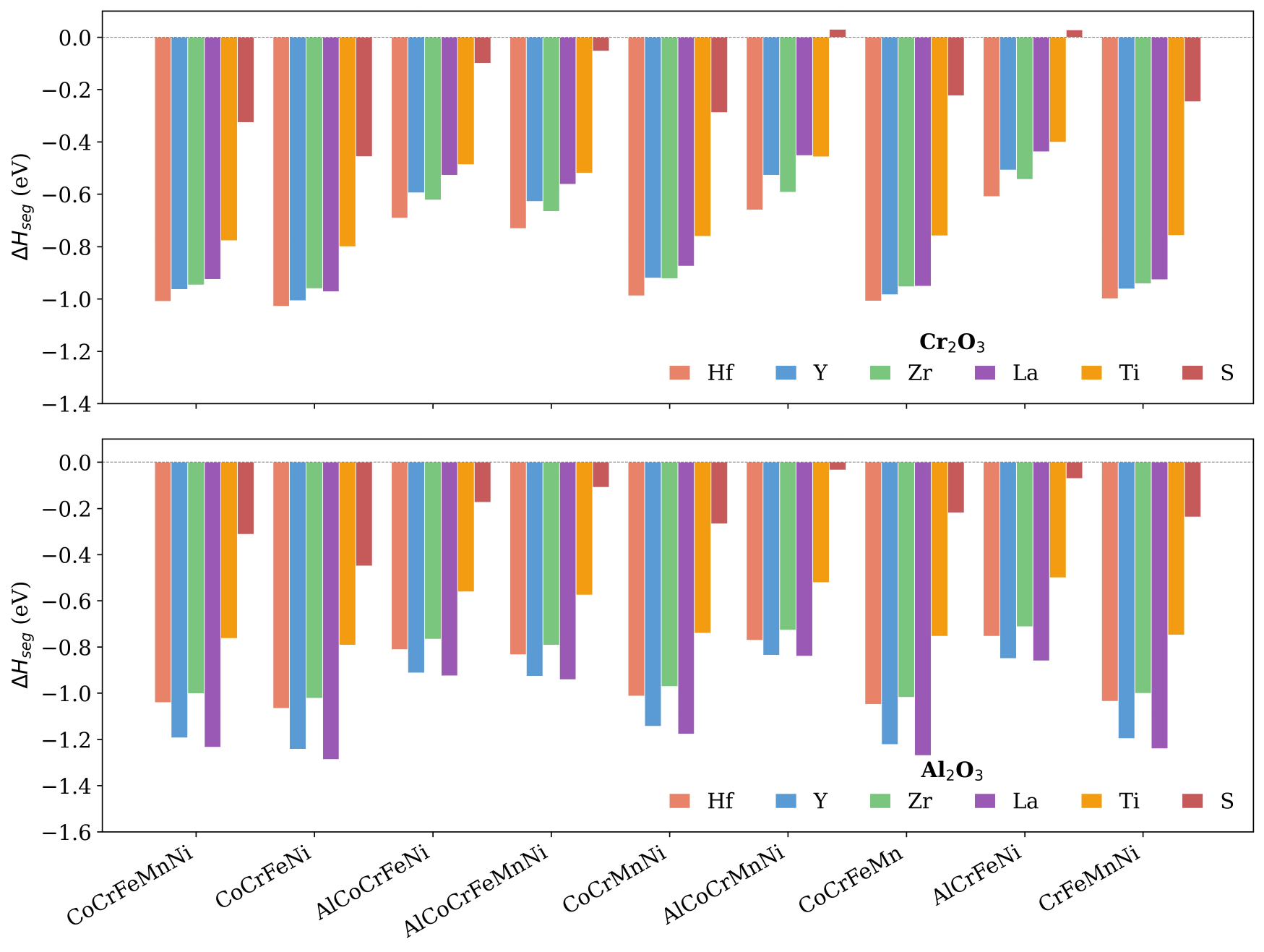}
	\caption{Segregation enthalpies of Hf, Y, Zr, La, Ti, and S across ten HEA families at (a)~\ce{Cr2O3} and (b)~\ce{Al2O3} interfaces. A striking ranking inversion occurs between the two oxides: Hf dominates at \ce{Cr2O3}, while La dominates at \ce{Al2O3} in most families.}
	\label{fig:seg_enthalpies}
\end{figure}

The segregation enthalpies of all five RE candidates and S were computed across all nine families at both oxide interfaces (Fig.~\ref{fig:seg_enthalpies}). At the \ce{Cr2O3} interface, the ranking is remarkably consistent across all families: Hf $>$ Y $\approx$ Zr $>$ La $>$ Ti. Hf is the strongest segregant in every case, with enthalpies ranging from $-0.66$~eV (AlCoCrMnNi) to $-1.03$~eV (CoCrFeNi). This ordering reflects the balance between the solute--oxide attraction (Term~1) and the solute--bulk penalty (Term~3): although the Y--O interaction ($-953$~kJ/mol) is weaker than Hf--O ($-1113$~kJ/mol), Y's more favorable matrix interactions partially compensate, placing it second. La ranks fourth because its large molar volume ($V_\text{La} = 22.55$~cm$^3$/mol) creates an unfavorable mismatch contribution.

At the \ce{Al2O3} interface, a dramatic ranking inversion occurs: La becomes the strongest segregant in most families, with Y second. The La segregation enthalpy reaches $-1.29$~eV in CoCrFeNi/\ce{Al2O3}, compared to $-1.07$~eV for Hf. To elucidate the mechanism, Table~\ref{tab:term_decomposition} presents the term-by-term decomposition for Hf and La in CoCrFeNi at both oxides.

\begin{table}[htbp]
	\centering
	\caption{Term-by-term decomposition of segregation enthalpy (eV) for Hf and La in CoCrFeNi, following Eq.~\eqref{eq:seg_enthalpy}.}
	\label{tab:term_decomposition}
	\begin{tabular}{lcccccc}
		\toprule
		RE & Oxide & Term~1 & Term~2 & Term~3 & Term~4 & $\Delta H_\text{seg}$ \\
		\midrule
		Hf & \ce{Cr2O3} & $-1.45$ & $-0.32$ & $-0.28$ & $+0.12$ & $-1.03$ \\
		La & \ce{Cr2O3} & $-1.17$ & $-0.32$ & $-0.21$ & $+0.24$ & $-0.80$ \\
		Hf & \ce{Al2O3} & $-1.38$ & $-0.28$ & $-0.28$ & $+0.12$ & $-1.07$ \\
		La & \ce{Al2O3} & $-1.52$ & $-0.28$ & $-0.21$ & $+0.24$ & $-1.29$ \\
		\bottomrule
	\end{tabular}
\end{table}

The inversion originates in Term~1: La's interaction with the Al-dominated oxide ($-1.52$~eV) exceeds that of Hf ($-1.38$~eV) at \ce{Al2O3}, driven by the more exothermic La--Al interaction ($\Delta H_\text{LaAl}^\circ = -231$~kJ/mol vs.\ Hf--Al: $-215$~kJ/mol), while the reverse holds at \ce{Cr2O3} ($\Delta H_\text{LaCr}^\circ = +93$~kJ/mol). The mismatch penalty for La (Term~4: $+0.24$~eV) is constant across oxide types and insufficient to reverse the ranking at \ce{Al2O3}. This oxide-dependent inversion has not been previously reported and has direct design implications: Hf should be preferred for chromia-forming alloys, while La or Y should be preferred for alumina-forming compositions. The finding challenges the common practice of applying a single RE across different oxide-forming systems without regard for oxide chemistry.

This prediction finds indirect support in the experimental literature. Lu et al.~\cite{lu2020hf} achieved exceptional oxidation resistance in Y/Hf co-doped AlCoCrFeNi, an alumina-forming alloy consistent with our prediction that both Y and Hf are effective at \ce{Al2O3}, with segregation enthalpies of $-1.15$ and $-1.07$~eV respectively. However, the model suggests that La alone could have provided even stronger interfacial enrichment at \ce{Al2O3} ($-1.29$~eV), a prediction that warrants experimental verification.

Ti consistently ranks weakest across all systems, with segregation enthalpies 30--50\% below the strongest segregant. Despite having a substantial Ti--O interaction ($-945$~kJ/mol), Ti's favorable matrix interactions reduce the net segregation driving force, consistent with the general observation that Ti additions are less effective than Hf or Y for adhesion improvement~\cite{han2025comparative}.

The S segregation enthalpy varies dramatically, from $-0.46$~eV (CoCrFeNi/\ce{Cr2O3}) to $+0.03$~eV (AlCoCrMnNi/\ce{Cr2O3}). In several Al- and Mn-containing alloys at \ce{Cr2O3}, S segregation becomes negligible or repulsive, a finding analyzed in Section~\ref{sec:mn_sulfur}.

\subsection{Sulfur vulnerability and RE thresholds}
\label{sec:s_vulnerability}
\label{sec:crossover}

\begin{figure}[!ht]
	\centering
	\includegraphics[width=1\linewidth]{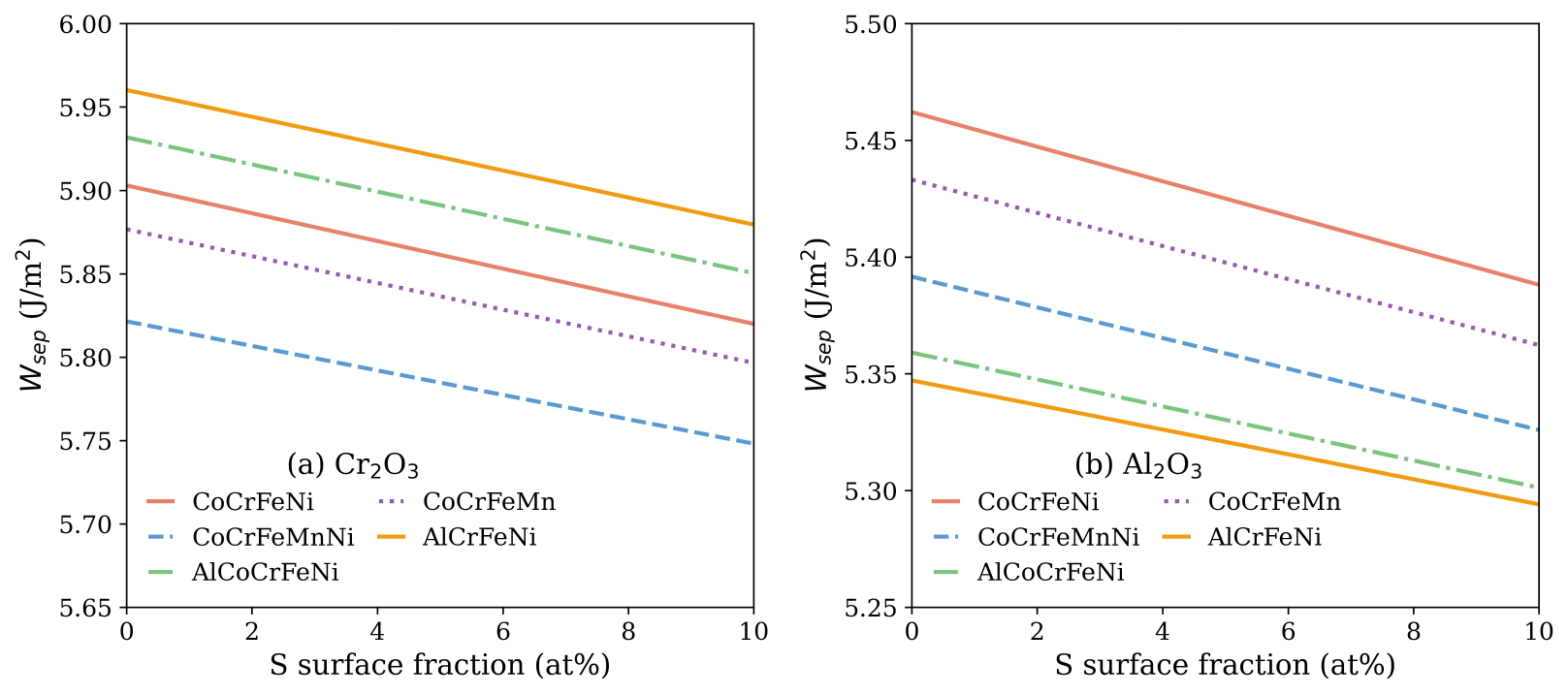}
	\caption{$W_\text{sep}$ as a function of S surface fraction for five representative alloys at (a)~\ce{Cr2O3} and (b)~\ce{Al2O3}. CoCrFeMn shows the steepest decline, while AlCrFeNi is least sensitive.}
	\label{fig:s_vulnerability}
\end{figure}

Given the central role of S in adhesion degradation, Fig.~\ref{fig:s_vulnerability} examines the sensitivity of $W_\text{sep}$ to S contamination for five representative families. All families show monotonic adhesion decline with increasing S concentration, but the rate and magnitude of decline differ substantially. CoCrFeNi and the Cantor alloy lose approximately 0.35~J/m$^2$ of adhesion at 10~at\% S surface fraction at \ce{Cr2O3}, while AlCoCrFeNi loses only 0.20~J/m$^2$. This reduced sensitivity of Al-containing alloys correlates with their weaker S segregation enthalpies: less S reaches the interface at equilibrium, and the interfacial bonding network is less disrupted per unit of S coverage.

The CoCrFeMn system shows the steepest decline in $W_\text{sep}$ per unit S at \ce{Cr2O3}, which might appear disadvantageous. However, as shown later in this section and in Section~\ref{sec:competitive_detail}, this system also has the lowest crossover concentrations for RE displacement of S, meaning that in practice, very little RE is needed to prevent S from reaching the interface.

\begin{figure}[!ht]
	\centering
	\includegraphics[width=1\linewidth]{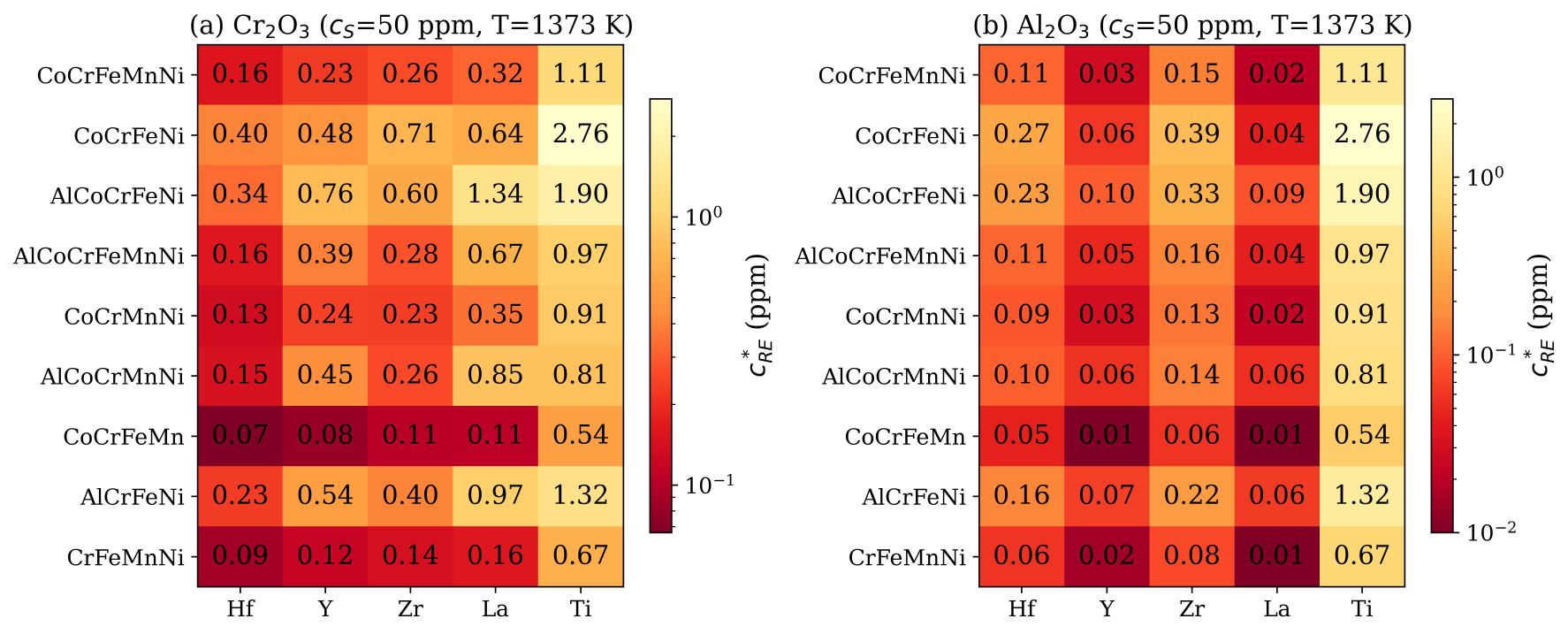}
	\caption{Crossover concentrations $c_\text{RE}^*$ (ppm) at 1373~K for each RE to displace 50~ppm S at (a)~\ce{Cr2O3} and (b)~\ce{Al2O3}. Log-scale color map highlights the two-order-of-magnitude range. CoCrFeMn requires only 0.07~ppm Hf at \ce{Cr2O3}.}
	\label{fig:crossover}
\end{figure}

Figure~\ref{fig:crossover} presents the crossover concentrations $c_\text{RE}^*$ for each RE--family--oxide combination at 1373~K, providing the minimum RE bulk concentration needed to achieve equal coverage with 50~ppm S. The crossover concentrations span more than two orders of magnitude, from 0.01~ppm (La at CoCrFeMn/\ce{Al2O3}) to 2.76~ppm (Ti at CoCrFeNi/\ce{Cr2O3}). Several significant patterns emerge.

First, all crossover concentrations for multicomponent HEAs are below 3~ppm at both interfaces, with the majority below 1~ppm. Since typical RE alloying additions are 0.05--0.5~wt\% (50--5000~ppm atomic), this confirms that practical RE levels exceed the thermodynamic minimum by 2--3 orders of magnitude. The excess serves primarily to compensate for kinetic limitations including slow bulk diffusion of RE to the interface, consumption by internal oxidation, and RE sequestration at grain boundaries~\cite{pint1996experimental,fritscher2023reactive}. The present predictions provide the thermodynamic lower bound, below which no amount of kinetic optimization can maintain a clean interface.

Second, CoCrFeMn consistently exhibits the lowest crossovers across all REs: 0.07~ppm Hf, 0.08~ppm Y, and 0.11~ppm Zr at \ce{Cr2O3}, approximately 6$\times$ lower than the corresponding CoCrFeNi values. The absence of Ni, which competes with REs for interfacial sites, removes a key impediment to RE segregation. This prediction identifies Ni-free chromia-forming HEAs that may require substantially lower RE additions for equivalent interfacial protection. Moreover, in our previous work, DFT calculations demonstrated that Ni sites are the most unfavorable for REs to segregate, further highlighting the effect of composition on segregation tendency~\cite{boakye2024reactive}.

Third, the oxide-dependent RE ranking from Section~\ref{sec:re_ranking} manifests clearly. At \ce{Cr2O3}, Hf requires the lowest concentration in most families. At \ce{Al2O3}, La dominates with values as low as 0.02~ppm in CoCrFeMnNi. The practical implication is stark: 0.02~ppm La provides the same interfacial protection at \ce{Al2O3} as 0.11~ppm Hf, a 5$\times$ advantage.	

\subsection{Competitive segregation: best versus worst cases}
\label{sec:competitive_detail}

\begin{figure}[!ht]
	\centering
	\includegraphics[width=0.7\linewidth]{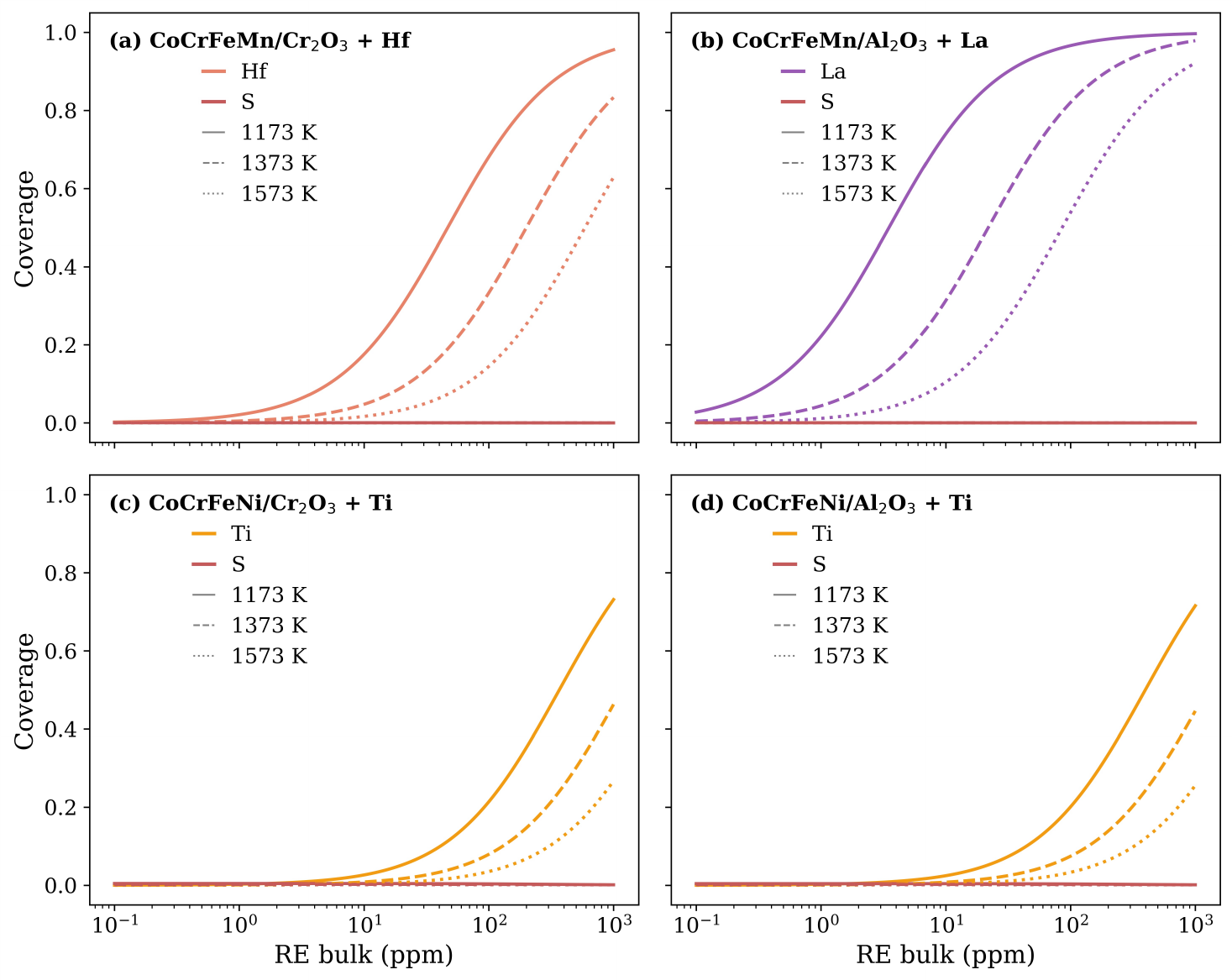}
	\caption{Competitive segregation for (a)~easiest case: CoCrFeMn/\ce{Cr2O3} + Hf ($c^* = 0.07$~ppm), (b)~easiest at \ce{Al2O3}: CoCrFeMn + La ($c^* = 0.007$~ppm), (c)~hardest case: CoCrFeNi/\ce{Cr2O3} + Ti ($c^* = 2.76$~ppm), and (d)~CoCrFeNi/\ce{Al2O3} + Ti ($c^* = 2.76$~ppm). CoCrFeNi + Ti represents the hardest combination because CoCrFeNi has the strongest S segregation ($-0.46$~eV) and Ti is the weakest RE.}
	\label{fig:competitive_detail}
\end{figure}

To illustrate the practical significance of the crossover thresholds, Fig.~\ref{fig:competitive_detail} compares the competitive segregation behavior of the easiest (CoCrFeMn + Hf at \ce{Cr2O3}, CoCrFeMn + La at \ce{Al2O3}) and hardest (CoCrFeNi + Ti at both oxides) alloy--RE--oxide combinations at three temperatures.	

In CoCrFeMn/\ce{Cr2O3} + Hf (Fig.~\ref{fig:competitive_detail}a), 1~ppm Hf achieves $>$95\% coverage at 1373~K while S coverage drops to $<$1\%. Even at 1573~K, the crossover occurs below 1~ppm, indicating robust protection across a wide temperature range. CoCrFeMn is the easiest alloy to protect because its strong Mn--S bulk interaction weakens S segregation ($\Delta H_\text{seg}^\text{S} = -0.22$~eV), while Hf maintains strong interfacial segregation ($-1.01$~eV). At \ce{Al2O3} (Fig.~\ref{fig:competitive_detail}b), CoCrFeMn + La achieves complete S displacement at even lower RE concentrations ($c^* = 0.007$~ppm), reflecting the oxide-dependent ranking inversion where La surpasses Hf at alumina interfaces.

In the hardest cases (Figs.~\ref{fig:competitive_detail}c,d), CoCrFeNi paired with Ti (the weakest RE) at both oxides, the crossover shifts to $\sim$2.8~ppm. CoCrFeNi has the strongest S segregation among all families ($-0.46$~eV at \ce{Cr2O3}), and Ti's relatively weak segregation ($-0.80$~eV) provides the smallest thermodynamic advantage over S. Nevertheless, even in this least favorable combination, $<$3~ppm Ti achieves majority coverage at 1373~K, confirming that the reactive element effect operates effectively across all multicomponent HEAs in this family.

\subsection{Temperature-dependent protection windows}
\label{sec:temp_crossover}

\begin{figure}[!ht]
	\centering
	\includegraphics[width=1\linewidth]{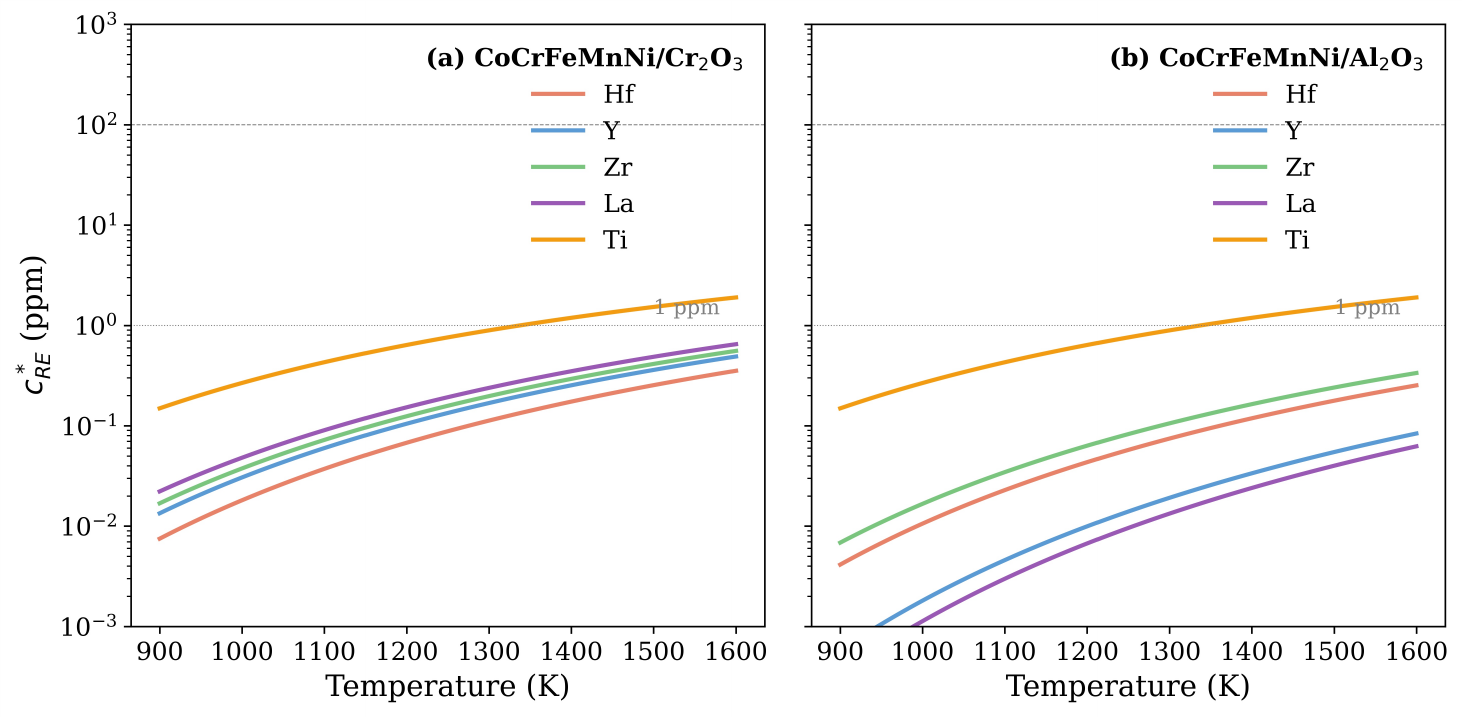}
	\caption{Temperature-dependent crossover concentration for all five REs to displace 50~ppm S in the Cantor alloy at (a)~\ce{Cr2O3} and (b)~\ce{Al2O3}. Horizontal lines mark 1~ppm and 100~ppm reference levels. All REs except Ti remain below 1~ppm at \ce{Cr2O3} up to 1500~K.}
	\label{fig:crossover_T}
\end{figure}

For practical alloy design, the crossover concentration must be known not just at a single temperature but across the operating range. Figure~\ref{fig:crossover_T} presents the crossover concentration as a continuous function of temperature for all five REs in the Cantor alloy.

The crossover concentrations increase monotonically with temperature, as expected from Eq.~\eqref{eq:crossover}, where higher temperatures reduce the Boltzmann selectivity between RE and S. At \ce{Cr2O3} (Fig.~\ref{fig:crossover_T}a), Hf remains below 1~ppm up to $\sim$1500~K, while Ti crosses the 1~ppm threshold at $\sim$1100~K. At \ce{Al2O3} (Fig.~\ref{fig:crossover_T}b), La maintains sub-ppm thresholds up to $\sim$1400~K, providing the broadest temperature window.

These curves define the RE protection window: the temperature range over which a given RE bulk concentration exceeds the crossover threshold. For example, 10~ppm Hf in the Cantor alloy provides thermodynamic protection at \ce{Cr2O3} up to approximately 1600~K, which is well above typical operating temperatures. The curves also reveal that Ti, while the weakest RE by segregation enthalpy, can still provide protection up to $\sim$1300~K at 10~ppm, which may be sufficient for moderate-temperature applications where Ti is preferred for other metallurgical reasons.

\begin{figure}[!ht]
	\centering
	\includegraphics[width=0.85\linewidth]{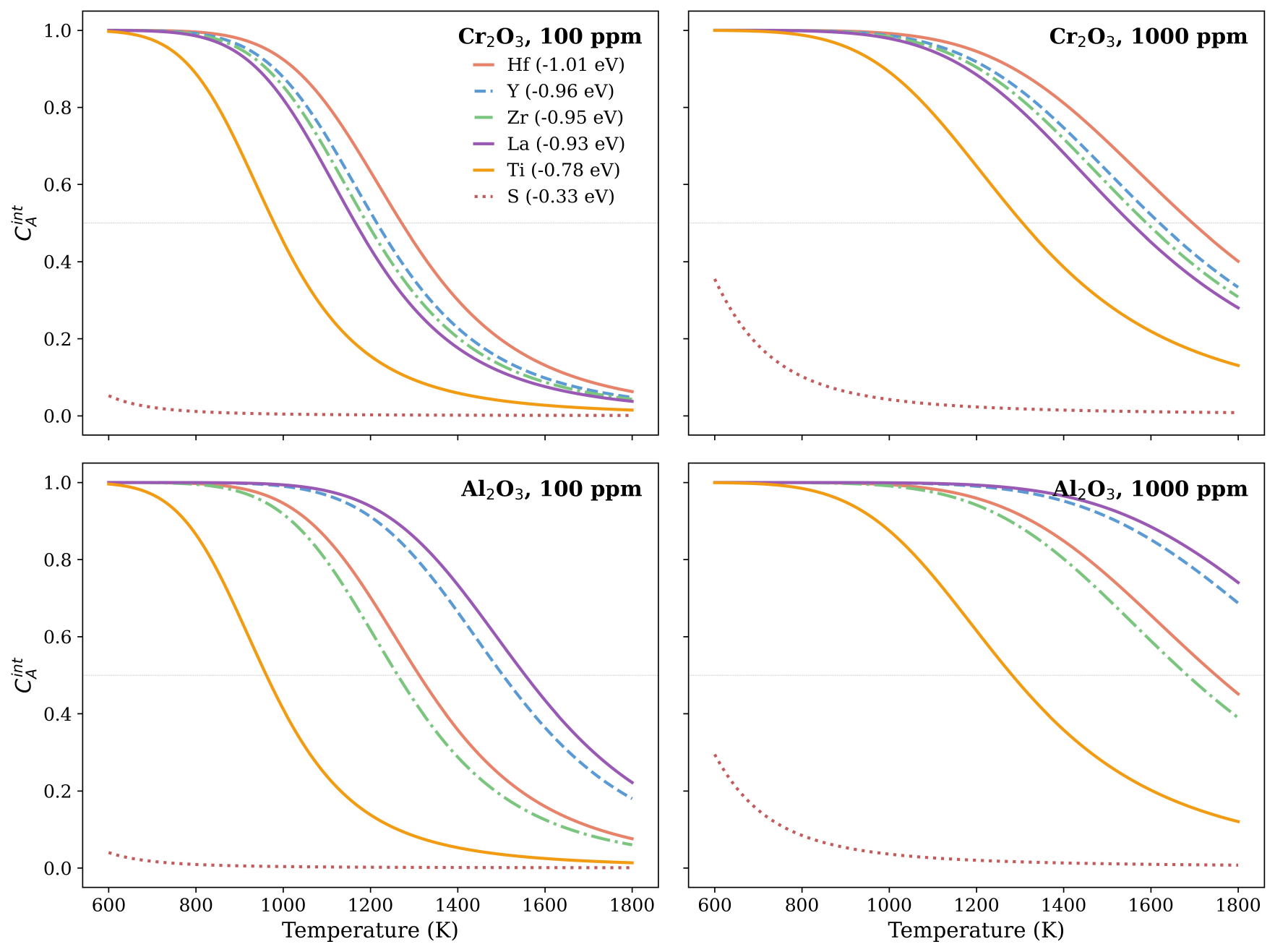}
	\caption{McLean isotherm predictions for the Cantor alloy at \ce{Cr2O3} (top) and \ce{Al2O3} (bottom) at 100~ppm (left) and 1000~ppm (right). All REs achieve near-saturation below 1000~K; S coverage remains low under all conditions.}
	\label{fig:mclean}
\end{figure}

Figure~\ref{fig:mclean} presents the McLean isotherm predictions for the Cantor alloy at both interfaces and two bulk concentrations. At 100~ppm, all five REs achieve near-saturation coverage below $\sim$1000~K at both interfaces. The desegregation transition occurs over a range of $\sim$400--600~K, with Hf maintaining the highest coverage at elevated temperatures at \ce{Cr2O3} and La at \ce{Al2O3}, consistent with the ranking inversion. At 1373~K, RE coverages range from 20\% (Ti) to 55\% (Hf) at \ce{Cr2O3}, while S coverage at the same concentration is below 5\%. This $>$10$\times$ disparity in coverage explains why ppm-level RE additions are effective: the exponential amplification from the Boltzmann factor overwhelmingly favors RE enrichment.

The \ce{Al2O3} results (bottom row) show broader saturation plateaus, with La maintaining $>$50\% coverage up to $\sim$1400~K at 100~ppm. Increasing to 1000~ppm extends all plateaus by $\sim$200~K, providing a practical buffer for extreme-temperature applications.

\subsection{Adhesion phase maps for composition optimization}
\label{sec:adhesion_maps}

\begin{figure}[!ht]
	\centering
	\includegraphics[width=1\linewidth]{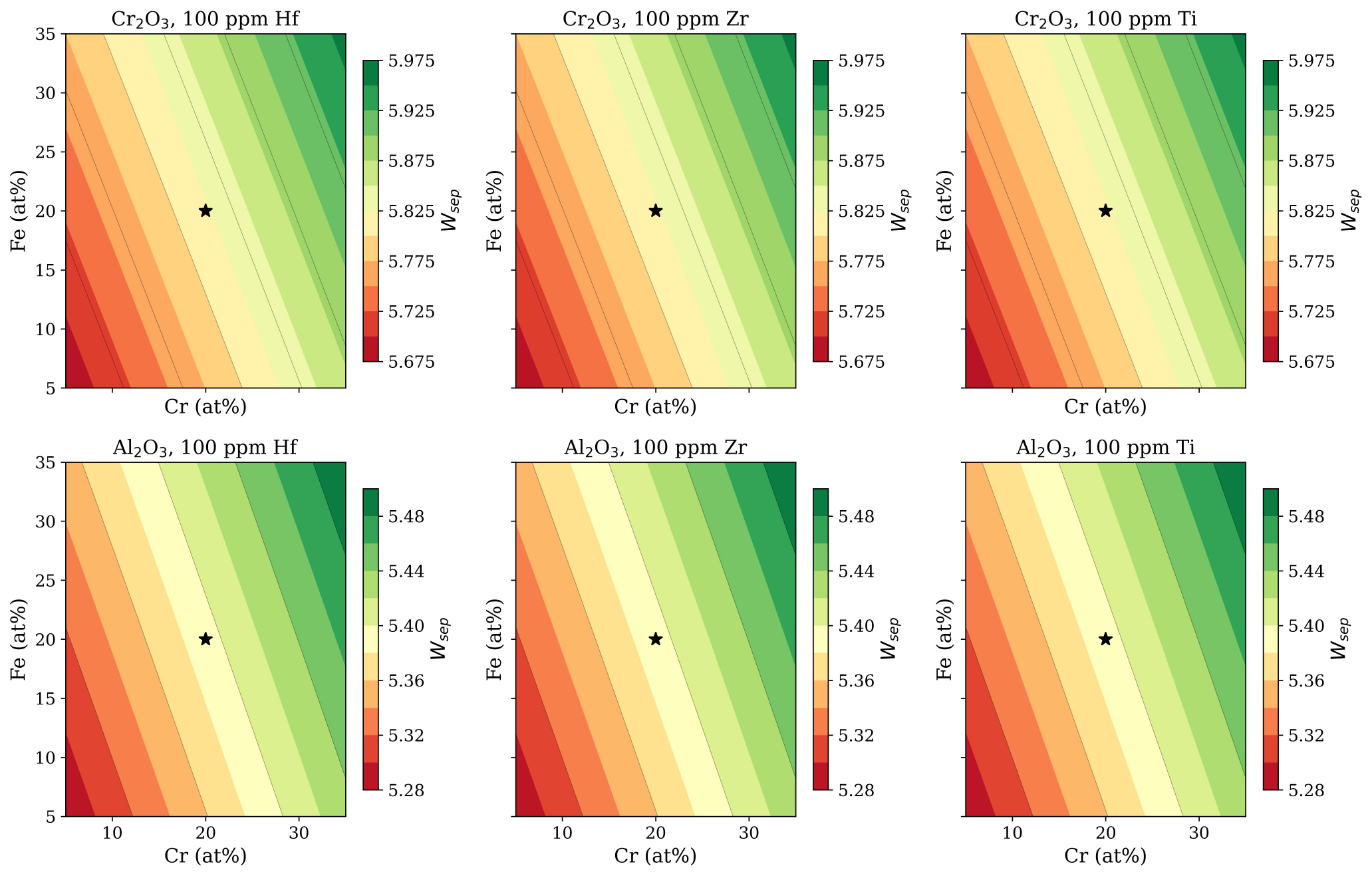}
	\caption{2D adhesion maps over (Cr,\,Fe) space for CoCrFeMnNi + 100~ppm Hf, Zr, Ti at \ce{Cr2O3} (top) and \ce{Al2O3} (bottom). Cr is the dominant lever at \ce{Cr2O3}; both Cr and Fe contribute at \ce{Al2O3}. Star marks the equiatomic point.}
	\label{fig:adhesion_maps}
\end{figure}

Two-dimensional adhesion maps over the (Cr, Fe) composition space for CoCrFeMnNi + 100~ppm RE (Fig.~\ref{fig:adhesion_maps}) reveal that $W_\text{sep}$ at \ce{Cr2O3} increases monotonically with Cr content, with increasing Cr from 10 to 35~at\% raising adhesion by approximately 0.3~J/m$^2$. The RE choice produces a uniform vertical shift ($\sim$0.02~J/m$^2$) without altering the contour topology, confirming that composition and RE effects are approximately separable at trace concentrations. This separability means that alloy designers can first optimize the base composition for adhesion, then independently select the RE.

For \ce{Al2O3}, both Cr and Fe contribute positively with diagonally oriented contours. The equiatomic composition does not coincide with the adhesion maximum at either interface, indicating that non-equiatomic, Cr-rich variants could provide substantial adhesion improvements.

The dominance of Cr as the adhesion lever at \ce{Cr2O3} connects to the percolation threshold concept. Xie et al.~\cite{xie2021effects} showed that a minimum of $\sim$10--12~at\% Cr is required for continuous passive film formation in Fe--Cr and Ni--Cr systems. The adhesion maps reveal an analogous, though gradual, dependence for scale adhesion, with Cr-rich compositions providing progressively stronger oxide--metal bonding. This finding is consistent with Holcomb et al.'s~\cite{holcomb2015oxidation} observation that increasing Cr content relative to Mn improved oxidation resistance in their series of CoCrFeMnNi variants. It also aligns with Butler and Weaver's~\cite{butler2016oxidation} finding that \ce{Cr2O3} constitutes the primary outer oxide layer in Al$_x$CoCrFeNi alloys with low Al content, where Cr dominates the interfacial chemistry.

\subsection{Adhesion enhancement and the distinction between segregation and bonding}
\label{sec:re_enhancement}

\begin{figure}[!ht]
	\centering
	\includegraphics[width=1\linewidth]{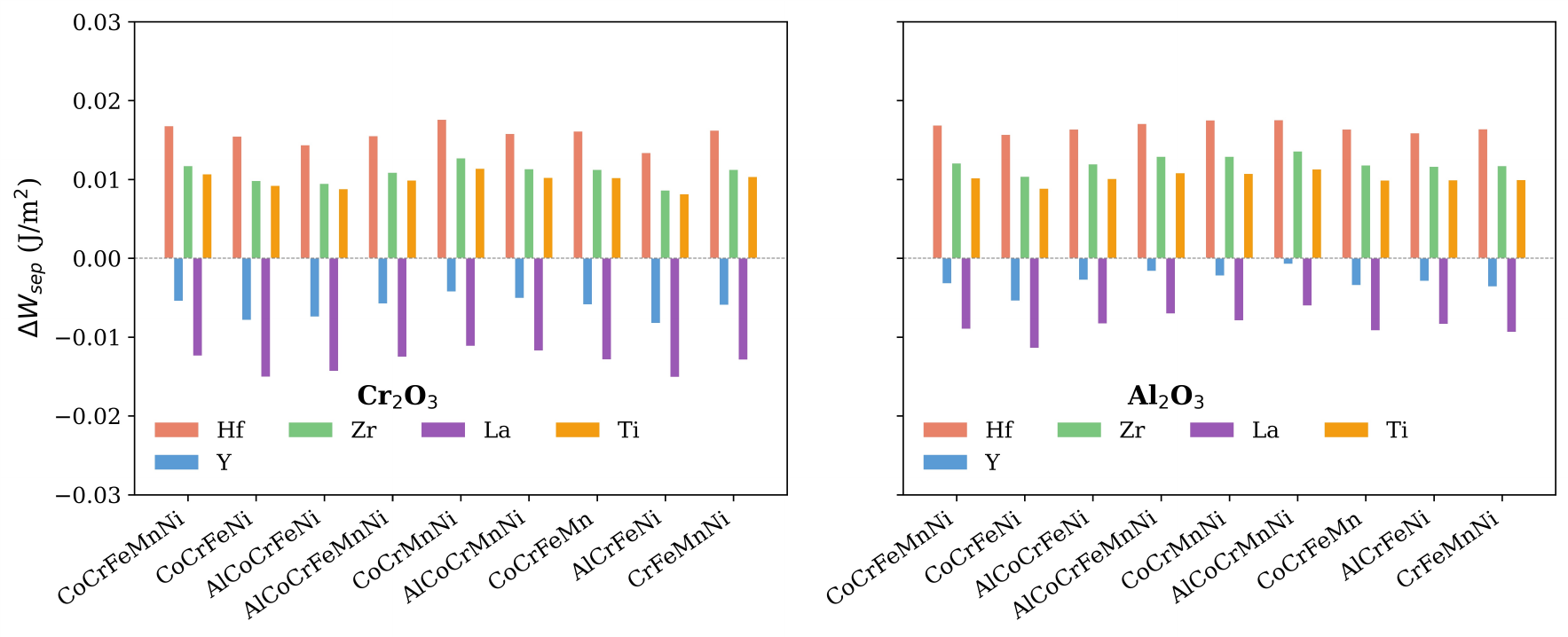}
	\caption{$\Delta W_\text{sep}$ at 1~at\% RE across all families at (a)~\ce{Cr2O3} and (b)~\ce{Al2O3}. Hf and Zr produce the largest enhancements.}
	\label{fig:delta_wsep}
\end{figure}

The adhesion enhancement $\Delta W_\text{sep}$ at 1~at\% RE (Fig.~\ref{fig:delta_wsep}) reveals an important subtlety: Hf consistently produces the largest $\Delta W_\text{sep}$ at \ce{Cr2O3}, followed by Zr---an ordering that differs from the segregation ranking where Y is nearly equal to Hf. This distinction arises because $\Delta W_\text{sep}$ depends on the RE--O interaction enthalpy once the RE reaches the interface (Hf--O: $-1113$~kJ/mol $>$ Zr--O: $-1098$ $>$ Y--O: $-953$~kJ/mol), not on the thermodynamic driving force for getting there. The element that segregates most readily is not necessarily the one that strengthens the interface most.

Segregation tendency and adhesion enhancement are two distinct properties that do not always correlate. To provide a single-metric ranking, we define a dimensionless composite RE effectiveness index that combines both:
\begin{equation}
	\eta_\text{RE} = \frac{|\Delta H_\text{seg}^\text{RE}|}{|\Delta H_\text{seg}^\text{S}|} \times \frac{|\Delta H^\circ_\text{RE--O}|}{|\overline{\Delta H}^\circ_\text{M--O}|},
	\label{eq:eta}
\end{equation}
where $|\overline{\Delta H}^\circ_\text{M--O}| = \sum_i C_i^S |\Delta H^\circ_{i\text{--O}}|$ is the surface-fraction-weighted average matrix--oxygen interaction enthalpy. The first ratio measures how much more strongly the RE segregates than S; the second measures how much stronger the RE--O bond is relative to the average matrix--O bond it replaces. Both ratios are dimensionless, and $\eta > 1$ indicates that the RE simultaneously out-segregates S and strengthens the interface beyond the baseline matrix--O bonding.

\begin{figure}[!ht]
	\centering
	\includegraphics[width=1\linewidth]{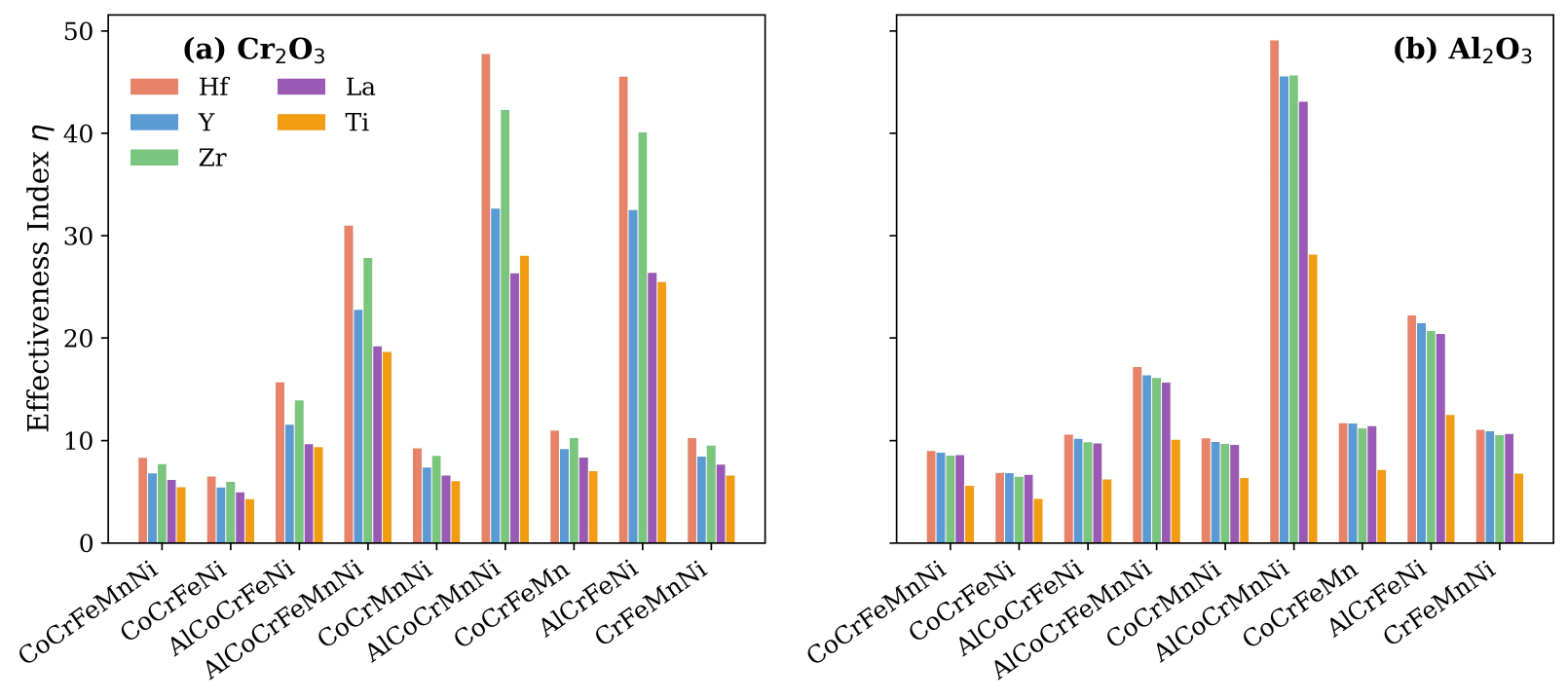}
	\caption{Dimensionless RE effectiveness index $\eta$ across all families at (a)~\ce{Cr2O3} and (b)~\ce{Al2O3}. Higher values indicate more effective REs. $\eta > 1$ means the RE both out-segregates S and forms stronger interfacial bonds than the matrix average. Hf dominates at \ce{Cr2O3}; La and Y dominate at \ce{Al2O3}.}
	\label{fig:effectiveness}
\end{figure}

Figure~\ref{fig:effectiveness} confirms the oxide-dependent inversion in a single dimensionless metric. At both interfaces, Hf ranks first across most families, with $\eta$ values exceeding 5 in CoCrFeNi to nearly 50 in AlCoCrMnNi, reflecting both its strong segregation ($-1.03$~eV) and its exceptionally strong Hf--O bond ($-1113$~kJ/mol, roughly 3$\times$ the matrix average). This is followed by Zr at the \ce{Cr2O3} and Y at the \ce{Al2O3} interfaces. The Al-containing alloys at \ce{Cr2O3} exhibit elevated $\eta$ values because their near-zero S segregation in the denominator amplifies the index, reflecting the physical reality that these alloys require minimal RE intervention for S protection. Consistent with the findings of Hou and Stringer~\cite{hou1995effect}, the RE effect exerts a stronger influence on chromia formers than on their alumina-forming counterparts. Hence, the effectiveness index provides a practical tool for RE selection: given an alloy composition and target oxide, the RE with the highest $\eta$ offers the best combination of segregation reliability and adhesion benefit.

\subsection{Effect of Mn on adhesion and sulfur resistance}
\label{sec:mn_sulfur}

\begin{figure}[!ht]
	\centering
	\includegraphics[width=0.8\textwidth]{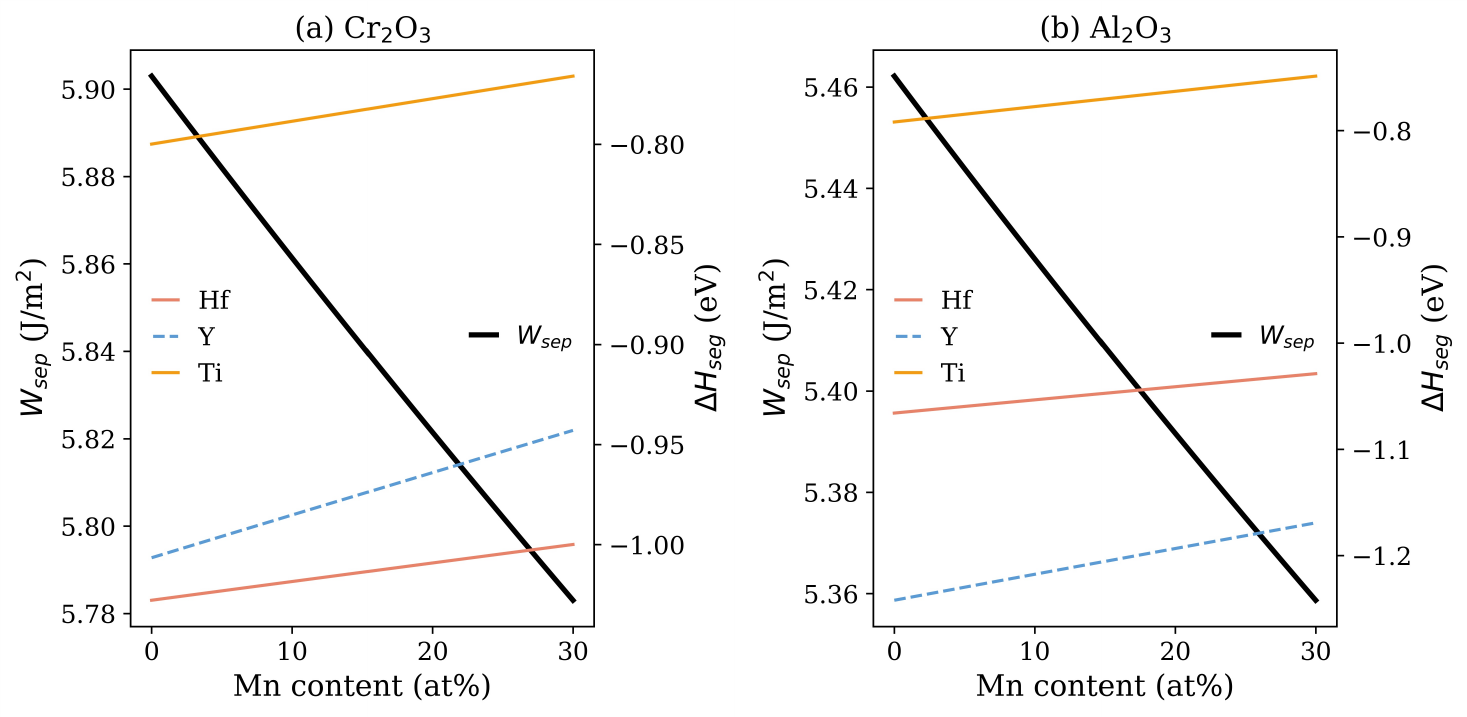}
	\caption{Effect of Mn content (0--30~at\%) on $W_\text{sep}$ (black, left axis) and RE segregation enthalpy (colored, right axis) for CoCrFe$_x$Mn$_y$Ni at (a)~\ce{Cr2O3} and (b)~\ce{Al2O3}. Mn reduces baseline adhesion by $\sim$0.1~J/m$^2$ over the full range.}
	\label{fig:mn_effect}
\end{figure}

The effect of Mn content on adhesion and RE segregation (Fig.~\ref{fig:mn_effect}) reveals a subtle trade-off. Increasing Mn from 0 to 30~at\% reduces $W_\text{sep}$ by approximately 0.1~J/m$^2$ at both interfaces, reflecting the dilution of stronger metal--O bonds by the weaker Mn--O interaction. This prediction aligns with the experimentally observed inferior scale adhesion of Mn-containing HEAs discussed in Section~\ref{sec:baseline_adhesion}.

However, the more significant finding concerns S resistance. Several Mn-containing alloys exhibit negligible or repulsive S segregation at \ce{Cr2O3}. The physical origin is the Mn--S interaction enthalpy ($\Delta H_\text{MnS}^\circ = -225$~kJ/mol), which is more exothermic than Ni--S ($-135$), Co--S ($-153$), or Cr--S ($-183$~kJ/mol). This strong Mn--S bulk interaction creates a thermodynamic trap that stabilizes S in the alloy matrix and penalizes its migration to the interface. Mn thus presents a double-edged character where it weakens the oxide--metal bond while simultaneously protecting the interface from S poisoning. Whether the net effect is beneficial depends on alloy purity. For ultra-low-S alloys ($<$5~ppm), the adhesion penalty dominates, and Mn is detrimental. For typical industrial alloys with 10--50~ppm S, the S-gettering effect may partially or fully compensate.

Dehury et al.~\cite{dehury2025experimental} provided direct experimental evidence for this duality. Their vacuum-annealing experiment at 1000$^\circ$C confirmed that Mn migration to the surface is oxidation-driven rather than thermally driven; in the absence of oxygen, no elemental segregation was observed. In oxidizing atmospheres, however, Mn migrates continuously to the surface and forms poorly adhering oxide phases. The present thermodynamic analysis explains why: the Mn--O interaction provides the driving force for outward diffusion, while the weak Mn--O bond at the interface ensures that the resulting scale is poorly anchored.

Importantly, the RE segregation enthalpies decrease only modestly with Mn content (5--10\% reduction from 0 to 30~at\% Mn), indicating that Mn does not impede the RE effect. The benefits of intrinsic S resistance and RE-mediated adhesion enhancement can coexist, suggesting that Mn-containing compositions like CoCrFeMn or CrFeMnNi may be effective hosts for RE doping requiring the lowest RE concentrations for S displacement while providing inherent S resistance as a secondary protection mechanism.

\subsection{Sulfur immunity phase diagram}
\label{sec:sulfur_immunity}

\begin{figure}[!ht]
	\centering
	\includegraphics[width=1\linewidth]{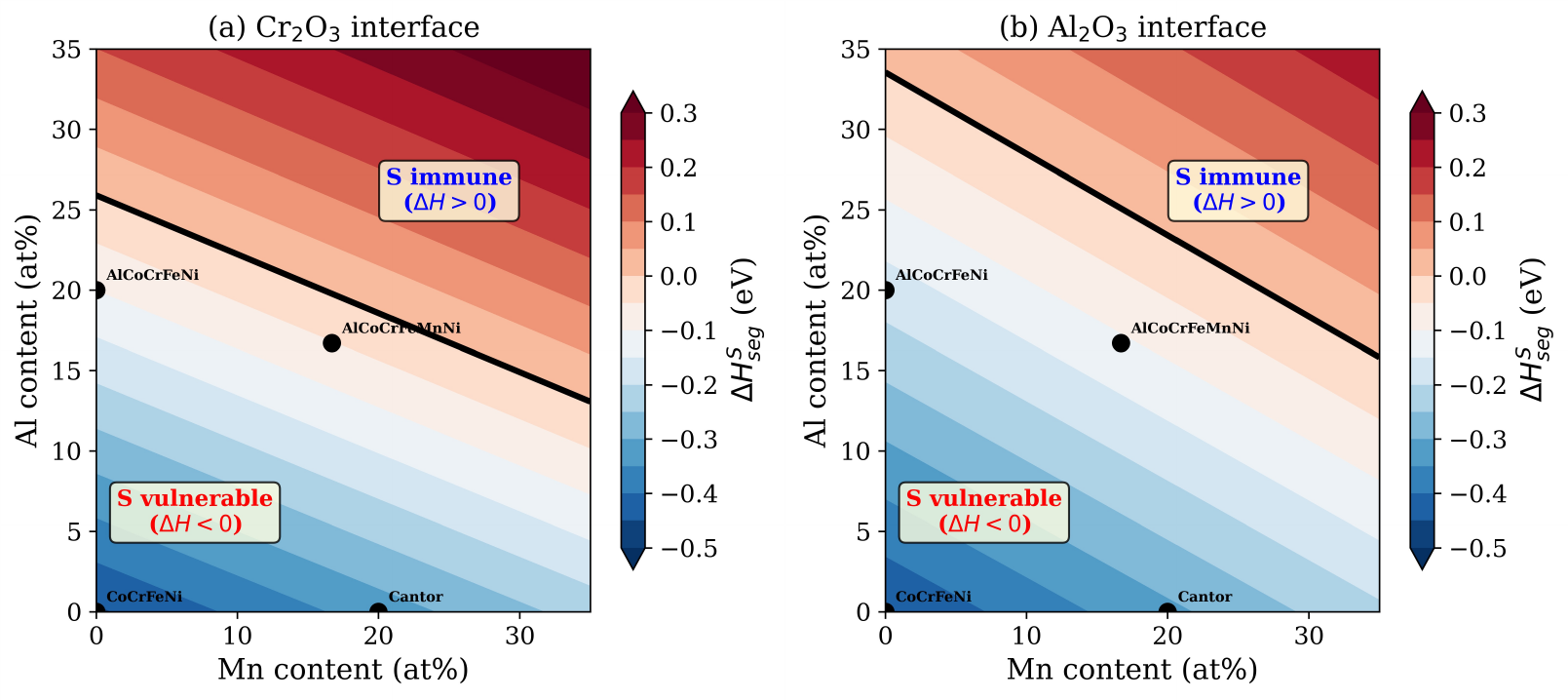}
	\caption{Sulfur immunity phase diagram in (Mn, Al) composition space at (a)~\ce{Cr2O3} and (b)~\ce{Al2O3} interfaces. The thick black contour marks $\Delta H_\text{seg}^\text{S} = 0$: above this line, S segregation is thermodynamically repulsive ($\Delta H_\text{seg}^\text{S} > 0$) and the alloy is inherently immune to S-induced adhesion loss. Below the line ($\Delta H_\text{seg}^\text{S} < 0$), S segregation is favorable and RE protection is required. Known alloy compositions are marked.}
	\label{fig:s_immunity}
\end{figure}

The observation that certain Al- and Mn-containing alloys exhibit near-zero or positive S segregation enthalpies motivates the construction of a sulfur immunity phase diagram. Figure~\ref{fig:s_immunity} presents $\Delta H_\text{seg}^\text{S}$ as a continuous function of Mn and Al content, with the balance shared equally among Co, Cr, Fe, and Ni. The contour $\Delta H_\text{seg}^\text{S} = 0$ defines the thermodynamic boundary above which S segregation becomes repulsive and the alloy is inherently immune to S-induced adhesion loss without any RE addition.

The immunity boundary follows an approximate diagonal: Al + Mn $\gtrsim$ 25~at\% at \ce{Cr2O3} and $\gtrsim$ 30~at\% at \ce{Al2O3}. This threshold arises from the combined S-trapping effect of Al ($\Delta H_\text{AlS}^\circ = -363$~kJ/mol) and Mn ($\Delta H_\text{MnS}^\circ = -225$~kJ/mol), both more exothermic than the Ni--S, Co--S, or Cr--S interactions.

The positions of known alloys on this map are revealing. CoCrFeNi sits deep in the S-vulnerable zone ($\Delta H_\text{seg}^\text{S} = -0.46$~eV at \ce{Cr2O3}), explaining why RE additions are critical for this system. The Cantor alloy (20\% Mn) approaches immunity but remains vulnerable. AlCoCrFeNi (20\% Al) lies near the boundary, consistent with its experimentally observed moderate S sensitivity. The senary AlCoCrFeMnNi is positioned on the cusp of immunity at \ce{Cr2O3}. This map constitutes, to our knowledge, the first thermodynamic phase diagram for sulfur vulnerability in HEAs and is directly testable by SIMS or AES measurements of interfacial S concentration in alloys positioned on opposite sides of the boundary.

\subsection{Universal scaling law for RE selection}
\label{sec:universal}

\begin{figure}[!ht]
	\centering
	\includegraphics[width=0.65\linewidth]{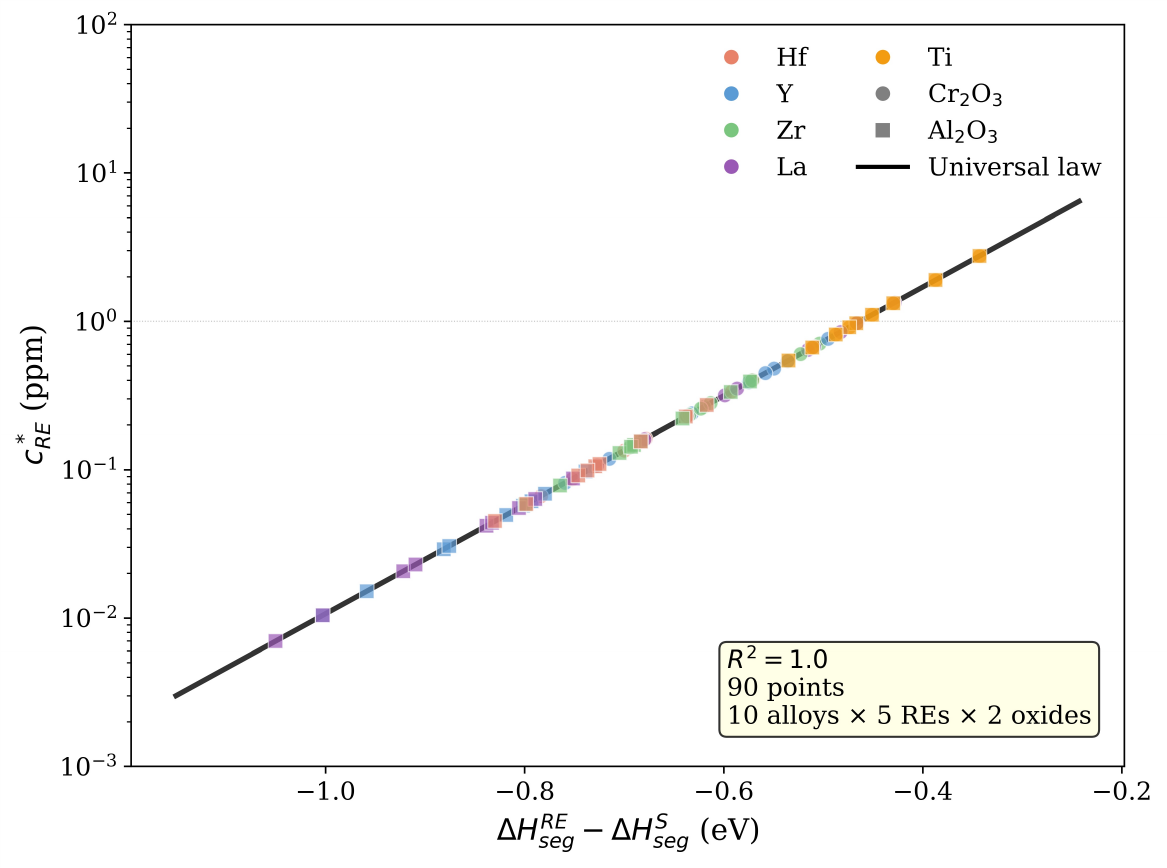}
	\caption{Universal scaling law: crossover concentration $c_\text{RE}^*$ versus segregation enthalpy difference $\Delta\Delta H = \Delta H_\text{seg}^\text{RE} - \Delta H_\text{seg}^\text{S}$ for alloy--RE--oxide combinations. All points collapse onto the single exponential predicted by Eq.~\eqref{eq:crossover}. Colors indicate RE types; shapes indicate oxide.}
	\label{fig:universal_scaling}
\end{figure}

Despite the apparent complexity of the crossover data spanning nine alloy families, five REs, and two oxide types, Eq.~\eqref{eq:crossover} predicts that all crossover concentrations should collapse onto a single master curve when plotted against the segregation enthalpy difference $\Delta\Delta H = \Delta H_\text{seg}^\text{RE} - \Delta H_\text{seg}^\text{S}$. Figure~\ref{fig:universal_scaling} confirms this prediction: all data points fall on the exponential master curve $c^* = c_\text{S} \exp(\Delta\Delta H / k_B T)$ with $R^2 = 1.000$. The entire complexity of alloy composition, RE choice, and oxide type reduces to a single thermodynamic parameter.

The practical consequence is that an alloy designer who knows $\Delta H_\text{seg}^\text{RE}$ and $\Delta H_\text{seg}^\text{S}$ for any system obtainable from a seconds-long MAM calculation can read the minimum RE concentration directly from the universal curve. The curve spans three orders of magnitude in $c^*$ over a 0.8~eV range in $\Delta\Delta H$, with a slope of approximately one decade per 0.27~eV at 1373~K. This law holds regardless of alloy composition, number of principal elements, RE identity, or oxide type.

\subsection{Inverse design of S-immune compositions}
\label{sec:inverse}

\begin{figure}[!ht]
	\centering
	\includegraphics[width=1\linewidth]{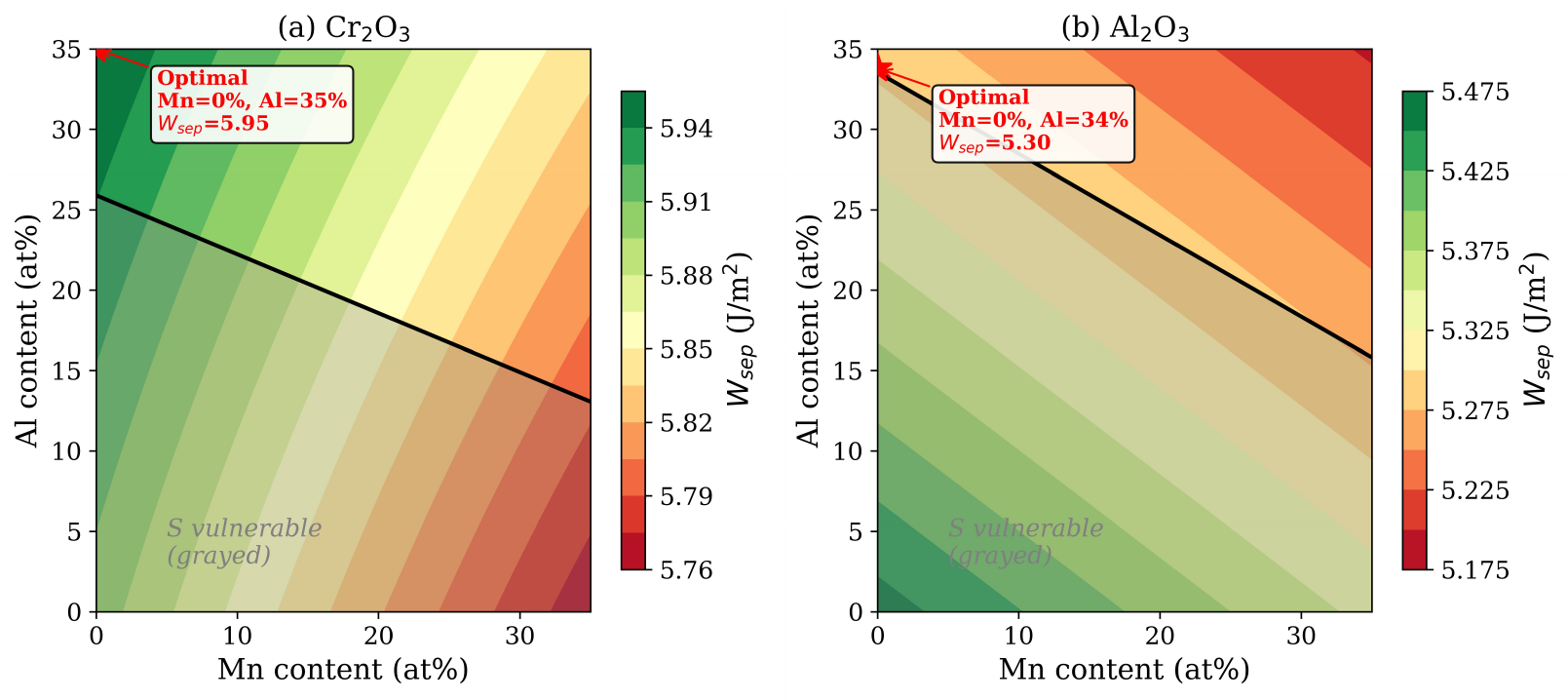}
	\caption{Inverse-designed optimal compositions: $W_\text{sep}$ contours overlaid with the S immunity boundary (thick black line) for (a)~\ce{Cr2O3} and (b)~\ce{Al2O3}. The S-vulnerable region is grayed. The red star marks the composition with maximum $W_\text{sep}$ within the S-immune zone: \ce{Co16Cr16Fe16Ni16Al35} with $W_\text{sep} = 5.95$~J/m$^2$ at \ce{Cr2O3}.}
	\label{fig:fig15optimal}
\end{figure}

The natural convergence of the S immunity boundary and the adhesion maps raises the question: what composition simultaneously achieves S immunity and maximum adhesion? Figure~\ref{fig:fig15optimal} overlays the $W_\text{sep}$ contours with the immunity boundary. The optimal S-immune composition for \ce{Cr2O3} is approximately \ce{Co16Cr16Fe16Ni16Al35} (at\%), with $W_\text{sep} = 5.95$~J/m$^2$, which is comparable to the best RE-enhanced adhesion in CoCrFeNi. This composition achieves strong adhesion without any RE addition, entirely through intrinsic thermodynamic resistance to S segregation. The high Al content creates both a strong Al--S bulk trap and favorable Al--O cross-interface bonding. For \ce{Al2O3}, the optimal is similar (\ce{Co17Cr17Fe17Ni17Al34}, $W_\text{sep} = 5.30$~J/m$^2$), offering dual-interface protection.

An important practical caveat must be noted. At 35~at\% Al, the alloy would likely adopt a BCC or B2 structure rather than the single-phase FCC that characterizes the Cantor family. Al$_x$CoCrFeNi alloys transition from FCC at $x < 0.3$ to FCC + BCC/B2 at $x \approx 0.5$ and to predominantly BCC at higher Al~\cite{butler2016oxidation}. While the present thermodynamic adhesion analysis is independent of crystal structure, practical implementation requires consideration of the associated mechanical property trade-offs. Non-equiatomic variants with moderately elevated Al (15--25~at\%) combined with 10--15~at\% Mn could potentially achieve S immunity while maintaining an FCC-based structure. The review by Veselkov et al.~\cite{veselkov2021high} concluded that AlCoCrFeNi alloys with additions of refractory elements represent the most promising direction for oxidation-resistant HEAs; the present analysis provides a quantitative thermodynamic basis for this recommendation.

\subsection{Connecting adhesion predictions to oxide scale morphology}
\label{sec:morphology}

The adhesion predictions can be related to the oxide scale morphologies reported in the experimental literature, providing additional validation. High-$W_\text{sep}$ alloys (CoCrFeNi, AlCoCrFeNi at \ce{Cr2O3}) form compact, continuous oxide scales with smooth metal--oxide interfaces~\cite{dehury2025experimental,lu2020hf}, because strong bonding resists delamination and maintains intimate contact during growth. Low-$W_\text{sep}$ alloys (Mn-rich compositions, Cr-free alloys) develop porous, fragmented scales with evidence of spallation and void formation~\cite{laplanche2016oxidation,holcomb2015oxidation,dehury2025experimental}.

The Kirkendall pores observed experimentally at the metal--oxide interface in Mn-containing alloys~\cite{laplanche2016oxidation,dehury2025experimental} represent a kinetic degradation mechanism that amplifies the weakness of thermodynamic adhesion. Vacancy condensation from preferential outward Mn diffusion creates voids that reduce the effective contact area, further decreasing practical adhesion below the thermodynamic $W_\text{sep}$. The combination of lower intrinsic adhesion and void-mediated area reduction explains the severe spallation observed in the Cantor alloy.

The experimental hierarchy also extends to aqueous environments. Chen et al.~\cite{chen2024selective} showed that selective oxidation in Al$_{0.1}$CrCoFeNi during corrosion alters the subsurface composition in a manner analogous to high-temperature oxidation: Cr, Co, and Fe preferentially oxidize, leaving a Ni-rich subsurface that hinders repassivation. The adhesion maps presented here (Fig.~\ref{fig:adhesion_maps}) provide a framework for predicting how such compositional perturbations shift $W_\text{sep}$ and would affect adhesion after multiple spallation--regrowth cycles.

\section{Conclusions}
\label{sec:conclusions}

High-throughput screening of oxide-scale adhesion across nine CoCrFeMnNiAl sub-families reveals an oxide-dependent RE ranking inversion: Hf dominates at \ce{Cr2O3} while La dominates at \ce{Al2O3}, driven by differences in RE--O and RE--matrix interaction enthalpies at the two oxide surfaces. This finding challenges the prevailing one-RE-fits-all approach and suggests that dual-RE strategies may be necessary for alloys forming both oxide types. All REs displace 50~ppm S with sub-ppm additions in most families, with practical RE levels exceeding these thermodynamic minima by orders of magnitude.

Mn plays a dual role: it weakens oxide--metal bonding by 0.06--0.08~J/m$^2$, consistent with the experimentally observed spallation tendency of the Cantor alloy relative to CoCrFeNi, while simultaneously providing intrinsic S resistance through strong Mn--S bulk interactions. A S immunity phase diagram identifies Al + Mn $\gtrsim$ 25~at\% as the compositional threshold above which S segregation becomes thermodynamically repulsive. All crossover concentrations collapse onto a single universal exponential governed by the segregation enthalpy difference $\Delta H_\text{seg}^\text{RE} - \Delta H_\text{seg}^\text{S}$, providing a transferable design rule for any HEA--oxide system. Inverse design identifies \ce{Co16Cr16Fe16Ni16Al35} as the optimal S-immune composition ($W_\text{sep} = 5.95$~J/m$^2$) without RE doping, though phase stability at high Al content constrains practical implementation. Cr content emerges as the dominant compositional lever for chromia adhesion, with non-equiatomic Cr-rich variants predicted to offer improvements of 0.15--0.30~J/m$^2$ over equiatomic compositions.

\section*{Acknowledgments}
This research was supported by the NSERC Alliance International Catalyst (ALLRP 592696-24), Canada, and the use of a high-performance computing system at the University of Manitoba and the Digital Research Alliance of Canada.

\section*{CRediT authorship contribution statement}
\textbf{Dennis Boakye}: Writing – review and editing, Writing – original draft, Visualization, Validation, Methodology, Investigation, Formal analysis, Data curation. \textbf{Chuang Deng}: Writing – review and editing, Supervision, Software, Resources, Project administration, Investigation, Funding acquisition, Conceptualization.

\section*{Declarations}
The authors declare that they have no known competing financial interests or personal relationships that could have influenced the work reported in this paper.

\section*{Data availability}
The Python code is available on reasonable request.

\section*{Supplementary information}
The supplementary information referenced in the main text is attached as the Supplemental material.

\scriptsize
\bibliographystyle{elsarticle-num}
\biboptions{sort&compress}
\bibliography{references}

\end{document}